\documentclass[manuscript,screen]{acmart}

\AtBeginDocument{%
  \providecommand\BibTeX{{%
    \normalfont B\kern-0.5em{\scshape i\kern-0.25em b}\kern-0.8em\TeX}}}

\setcopyright{cc}
\copyrightyear{2024}
\acmYear{2024}
\acmDOI{XXXXXXX.XXXXXXX}

\acmConference[CHI WS '24]{CHI2024 Workshop Designing (with) AI for well-being}{May 12, 2024}{Honolulu, Hawai’i}
\acmBooktitle{CHI2024 Workshop Designing (with) AI for well-being,
 May 12, 2024, Honolulu, Hawai’i} 
\acmISBN{978-1-4503-XXXX-X/18/06}

\usepackage{color}
\usepackage{colortbl, booktabs}
\usepackage{xcolor}
\usepackage{multirow}
\usepackage{subcaption}

\newif\ifdraft
 \drafttrue
 \draftfalse
\ifdraft
   \newcommand{\todo}[1]{\textsf{\textbf{\textcolor{red}{[TODO: #1]}}}}
  \newcommand{\key}[1]{\textsf{{\textit{\textcolor{red}{#1:}}}}}
  \newcommand{\yuri}[1]{\textsf{\textcolor{blue}{\textbf{Yuri:} \textit{#1}}}}
\else
  \newcommand{\todo}[1]{}
  \newcommand{\key}[1]{}
  \newcommand{\yuri}[1]{}
\fi

\begin{document}

\title{What Should Be Considered to Support well-being with AI: \\Considerations Based on Responsible Research and Innovation}

\author{Yuri Nakao}
\email{nakao.yuri@fujitsu.com}
\orcid{0000-0002-6813-9952}
\affiliation{
  \institution{Graduate School of Arts and Sciences, The University of Tokyo}
  \streetaddress{}
  \city{Meguro-ku}
  \state{Tokyo}
  \country{Japan}
  \postcode{153-8902}
}
\affiliation{%
  \institution{AI Trust Research Centre, Fujitsu Research, Fujitsu Limited}
  \streetaddress{}
  \city{Kawasaki}
  \state{Kanagawa}
  \country{Japan}
  \postcode{211-8588}
}


\begin{abstract}
To achieve people's well-being with AI systems, we should enable each user to be guided to a healthier lifestyle in a way that is appropriate for her or him. However, there is a dilemma between general well-being as defined in academic and medical discussions and the autonomy users should have when deciding how to promote their well-being. In this position paper, we discuss the difficulty for AI application developers to fully consider in the design phase what might happen to the user, taking an example of a running application (app). We sort out the required factors to enable AI apps that support well-being to address the dilemma between unilaterally defined well-being and human autonomy based on the four dimensions required for responsible innovation: inclusion, anticipation, reflexivity, and responsiveness. 
\end{abstract}






\maketitle

\section{Introduction}
Achieving people's well-being with AI systems requires that each user is guided to a healthier lifestyle in a way that is appropriate for her or him. Although well-being has diverse definitions~\cite{calvo2014positive}, leading a healthy lifestyle is one of the most representative aspects of well-being. A healthy lifestyle often varies from individual to individual and cannot be defined in a top-down manner. For example, while moderate exercise is important for almost everyone, how much exercise is needed and at what time of day varies from person to person. A habit that is easy for one person may be very difficult for another.  Habits that are too difficult do not lead to a mentally healthy lifestyle.

This ambiguity, where although the larger goal (i.e., a better lifestyle) is clear, it is unclear how individuals should do a specific activity to achieve that goal, makes support by AI for well-being not straightforward. Because people's behavior is always influenced by technology~\cite{Winner1980Do}, their thoughts about how they want to achieve well-being are also influenced by technology. At the same time, people also need to be able to make decisions with autonomy in order for their personal preferences and their state of mind to be appropriately reflected in AI systems. Both the importance of well-being and human autonomy have been argued in the discussions related to socially responsible AI systems. For example, Floridi et al.~\cite{floridi2018ai4people} discussed the need for human autonomy while they included promoting well-being in the ethical principles of AI. In the ethically aligned design that prioritizes human well-being~\cite{ieee2017ethically}, it has also been suggested that human autonomy could be influenced by autonomous systems. There remains a dilemma between general well-being as defined in academic and medical discussions and the autonomy that individuals should have in the way they promote their own well-being.

To overcome this dilemma and create AI systems that support human well-being, this position paper discusses what to consider when designing AI applications for people's well-being by examining the characteristics of modern AI and referring to a research and development governance methodology called Responsible Research and Innovation (RRI).

\section{How well-being should be promoted by technology: An Example}

To consider what problems might arise when trying to encourage well-being with AI applications (apps), we take a running app as an example of a system that includes AI technology. That app sends you a push notification if you have not run in a while, asking if you want to run this month, and it incorporates gamification, such as earning badges based on distance, speed, and frequency of running. However, just being motivated to run long distances, in short periods of time, and at high frequency can lead some users to put a great deal of stress on their own legs, which may lead to leg injuries and make it more difficult to run than before\footnote{This actually happened to the author of this paper. As a result, the author received rehabilitation treatment at a hospital.}. This is a simple example of the perverse effect that apps with AI that intend to support well-being can have on a person's health.

To prevent this situation, users should be allowed to decide how they want their activities to be promoted more flexibly. In the field of philosophy of technology, Verbeek~\cite{verbeek2011moralizing} names the effect of technology on the extension of users' moral values as ``mediation.'' According to him, what kind of behavior, in what way a designed technology intends to promote, and the result of the mediation must be examined to execute a moral assessment of a technology~\cite{verbeek2011moralizing} (p.130). Following this discussion, we should examine what kind of effect an app will bring to a user's physical and mental health when evaluating the moral values the app promotes. Additionally, Verbeek also argues that users should be able to make themselves what they desire in the use of technology~\cite{verbeek2011moralizing} (p.84). Based on this, there needs to be a function that allows users to determine how the behaviors that lead to a healthy life can be amplified by technology in the use of the app. If there were a function in the app that allowed users to modify what they pursue as healthy living, problems such as the above example could have been prevented. The point is that in order for technology to promote well-being, there needs to be a part of its design that can reflect how each user wants to achieve her/his goal.

\section{The design difficulties based on the characteristics of AI technology} 
When considering how personal values should be mediated in AI systems, it is necessary to take into account the differences between other technologies and modern AI technologies, that is, AI technologies can be personalized for each user. Modern data-driven AI technologies change their responses based on the interaction with their users. The users' behavior after receiving a response from the AI technology will also change based on its response. Thus, recursive changes occur between AI and users. 

This recursive change makes it difficult for AI app developers to fully consider what might happen to the user in the design phase. 
Developers design how the AI app will change the information it presents to the user in response to the user's behavior. However, it is almost impossible to completely predict at the design stage how the system will actually change for months or years in the actual use of the app. For example, long-term use of an AI system might accelerate the same behavior of users. Users might overeat certain types of food or overdo certain types of exercise. If those things happen, as a consequence, the user may be less likely to improve his or her lifestyle due to injury, dietary bias, or boredom than if s/he did not use the AI system.

To address this difficulty, while the individual users use the AI app, it is necessary to make the AI app reflect the effect of technology on the users so that the system will promote their desired behavior. This can be achieved, for example, by adding functions to AI apps to change information presentation policies or by users committing to modifying the app by sharing their opinions on a specific website, as in the case of open-source software. In this way, AI apps to support well-being require a mechanism in which the value of the apps can be modified by the interaction between the technology and the user groups after it is implemented in society, rather than the designer deciding the value to be included in the technology at the design stage.




\section{How AI APPS should be designed: RRI-based considerations}
How should the design of the technology done by the developer prior to social implementation be so that it can be modified by the user groups after the AI apps are implemented in society? In this section, we sort out what should be considered during technology development with reference to Responsible Research and Innovation (RRI), one of the methodologies of governance for research and development.
RRI is a methodology on how to make research and development that leads to innovation socially responsible~\cite{STILGOE2013Developing}. In RRI, it has been proposed to divide factors required for socially responsible innovation into several elements and to integrate and implement various methods in a way that ties them to each element~\cite{STILGOE2013Developing}. While several different ways of dividing have been proposed~\cite{Fraaije2020Synthesizing,Owen2014UK,Foley2016Towards}, in the following, we will examine how the AI apps for well-being should be designed using the four most commonly used elements of inclusion, anticipation, reflexivity, and responsiveness, which are referred to as the four ``dimensions''~\cite{STILGOE2013Developing}.


\subsection{Inclusion}
This dimension implies including stakeholders of research and development, e.g., non-experts such as citizens, in innovation and its governance in order to legitimize the innovation process~\cite{STILGOE2013Developing}.  In the design of technology, including stakeholders in the design of applications through methods such as participatory design~\cite{bodker2009participatory}, co-design~\cite{Sanders2008CoCreation}, and value-sensitive design~\cite{friedman2019value} can be effective in introducing perspectives to innovation that cannot be obtained by the developers alone. Of course, stakeholder participation does not completely guarantee the correctness of the designed output~\cite{Sloane2022Participation}. Additionally, how to select the stakeholders to participate and what ways of participation are desirable should be carefully considered on an application-by-application basis. Nevertheless, considering what is likely to happen in the use of AI apps by the people involved will give an important perspective when developing AI apps that support well-being over time.

\subsection{Anticipation}
This dimension implies considering innovation risks and new innovation opportunities through thinking about possible futures in the process of research and development~\cite{STILGOE2013Developing}. In RRI, the future may be considered in the span of a few decades or a century.  On the other hand, when we consider the impact of an AI application on a single person, in many cases, it may be sufficient to consider the future in a few years or, at most, a few decades. Of course, this does not mean that we only need to consider the near future because it is possible that we also need to consider the potential impact over a longer span of time as to what will happen when many people use the app. At least, the future impact of the developed app on the well-being of individuals should be scrutinized in discussions with various stakeholders during the technology development process.

\subsection{Reflexivity}
This dimension implies that developers and researchers reflect on their own, often domain-dependent, assumptions and scrutinize the values that underlie technological development and its governance~\cite{STILGOE2013Developing}. Rather than individual developers' reflection on their own activities, the word reflexivity involves a collective reflection with non-experts and experts from other fields, such as researchers in the humanities and social sciences, to take into account the values of society and to review the developers' own values as a collective group of developers. When supporting well-being, it is necessary to question whether the policies assumed, e.g., the more exercise is promoted, the better, or a healthy diet consisting mainly of vegetables is preferred above all, as long as the weight is kept under control, are correct. For example, whether the lifestyle promoted by an AI app is a good one should be examined, especially when the developer is motivated to sell a particular sports equipment or a particular dietary supplement. While it is conventionally thought that such reflection should be done institutionally~\cite{Wynne1993Public}, it is also thought that it can be held collaboratively, with each person realizing the limits of what s/he is thinking in the course of discussions among those who participate in the design of the app~\cite{Grimpe2020From}.

\subsection{Responsivenesss}
This dimension means that the innovation process and its outputs can change in response to the values of stakeholders and society~\cite{STILGOE2013Developing}.   In the context of developing AI apps that support well-being, this includes the fact that the output from the apps and the policies regarding the information presented to the user are responsive to the opinions of the user groups. To achieve this responsiveness, it is necessary to consider what changes will need to be realized in the application.
It will also be necessary to identify the means by which the app can accept changes in the user groups after the app has been implemented in society. For example, there are several options, such as adding the functions to change the policy of information presentation to the app, or creating a mechanism that allows many people to commit to modifying the policy of the AI app referring to open source software platforms or Wikipedia.

\section{Conclusion}
In this position paper, we discussed what needs to be considered for designing AI to support well-being, taking cues from the philosophy of technology and RRI. People need to be able to decide at the design stage of an AI app what activities will be facilitated by the app and how.  On the other hand, it is not possible to identify all the effects of recursive changes between AI apps and users during the design phase. Therefore, it is important to make the policy of what information AI presents changeable by user groups after the app is in use in society. We examined what is needed to make this possible based on the four dimensions required for innovation at RRI: inclusion, anticipation, reflexivity, and responsiveness. Based on this discussion, we argue that it is necessary to work with diverse stakeholders to consider the possible impact of the application on individual users and society, taking into account social values outside the group of developers and design participants and how to respond to the opinions of the user group and its changes. 

Developing AI to support well-being based on the directions presented in this position paper will allow humans to autonomously specify the AI's behavior while AI and people are changing interactively. This would lead to the development of AI apps that support more human-oriented well-being, such as helping each person reach a better state of being in the way that is best for that person rather than a unilaterally defined well-being. This may also lead to a reduction in the number of people who suffer unexpected harm from well-being-enhancing AI applications, as I did when I injured my leg as a result of an AI application.

\begin{acks}
This work was supported by JST, ACT-X Grant Number JPMJAX21AJ, Japan.
\end{acks}

\bibliographystyle{ACM-Reference-Format}
\bibliography{sample-base}










\end{document}